\documentclass[%
preprint,
amsmath,amssymb,
aps,
prb,
floatfix,
]{revtex4-1}
\usepackage{graphicx}
\usepackage{hyperref}
\usepackage{color} 
\usepackage{array}
\newcolumntype{C}[1]{>{\centering\arraybackslash}p{#1}}
\begin{document}
	
	
\title{Exploring complex pattern formation with convolutional neural networks}
	
	\author{Christian Scholz}
	\email{christian.scholz@hhu.de}
	\author{Sandy Scholz}%
	\affiliation{%
		Institut f\"ur Theoretische Physik II: Weiche Materie, Heinrich-Heine-Universit\"at D\"usseldorf, 40225 D\"usseldorf, Germany
	}%
	
	\date{\today}
	
	\begin{abstract}
Many nonequilibrium systems, such as biochemical reactions and socioeconomic interactions, can be described by reaction-diffusion equations that demonstrate a wide variety of complex spatiotemporal patterns.
The diversity of the morphology of these patterns makes it difficult to classify them quantitatively and they are often described visually. Hence, searching through a large parameter space for    patterns is a tedious manual task. 
We discuss how convolutional neural networks can be used to scan the parameter space, investigate existing patterns in more detail, and aid in finding new groups of patterns. As an example, we consider the Gray-Scott model for which training data is easy to obtain. Due to the popularity of machine learning in many scientific fields, well maintained open source toolkits are available that make it easy to implement the methods we discuss  in advanced undergraduate and graduate computational physics projects.
	\end{abstract}
	
	\maketitle
	
	
\section{\label{sec:level1} Introduction}

Many systems in nature for which energy is constantly injected and then dissipated via internal degrees of freedom demonstrate complex patterns far from thermal equilibrium.~\cite{Cross2009} Nonlinear reaction-diffusion equations are a class of models that exhibit such complex behavior. An example is the Gray-Scott model,~\cite{Gray1983autocatalytic} which is represented by two coupled reaction-diffusion equations with cubic reaction terms that arise from autocatalysis, that is, the reactants activate or inhibit each other's creation.~\cite{Murray2003} Despite having only three parameters, the system can show many  different stationary and spatiotemporal solutions, such as spiral waves, moving or self-replicating spots, and phase turbulence.~\cite{Pearson1993,Mazin1996} Due to the nonlinearity of the model, it is not easy to investigate the system analytically and in many cases solutions have to be obtained  numerically. 

The solutions depend on the system parameters and on the initial conditions. No simple set of order parameters have been found that describe which type of pattern is observed for a specific set of the parameters and initial conditions. Often a visual scan of the parameter space is necessary, which is especially difficult in experimental realizations of chemical reaction-diffusion systems, for which a continuous influx of chemicals has to be provided for hours or even days to scan through multiple parameters.~\cite{Castets1990, Lee1993, Ouyang1991}

The  problem is that there is no obvious way to classify and quantify the occurring patterns, except for cases that display only a few types of clearly distinguishable patterns.~\cite{Mecke1996,Guiu2012} One  way to treat this problem, for which pattern classification is based on human perception, is to fit a logistic regression model of the probability that a pattern belongs to a certain class to the data. However, for two- and three-dimensional datasets, the  number of input variables is too large. For such problems convolutional neural networks (CNNs) are a particularly useful model for classification,~\cite{Bishop2006pattern, Monti2017geometric} and have become an essential tools in a growing number of scientific fields.

In the following, we demonstrate how CNNs can be used to explore  the parameter space of the Gray-Scott model and classify patterns. Our use of CNNs is suitable as an introduction to machine learning in advanced undergraduate or graduate computer physics courses and provides a new perspective on pattern formation in reaction-diffusion systems. The required code is written in Python and is available as supplementary material.~\cite{Scholz2021, Scholz2021b}

\begin{figure*}[t]
\includegraphics[width=\textwidth]{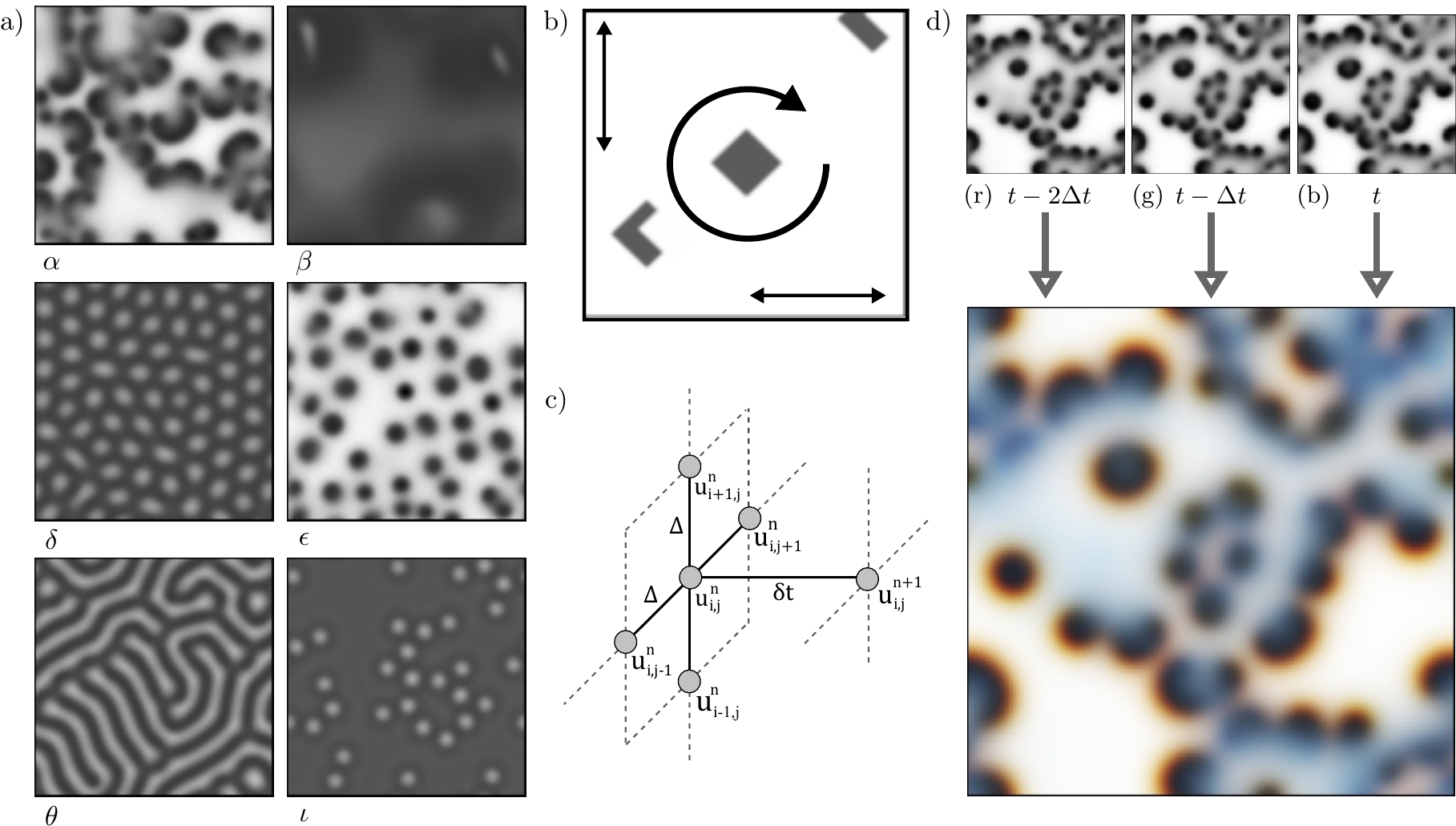}
\caption{\label{FigureClasses} (Color online) (a) Typical classes of heterogeneous and spatiotemporal patterns exhibited by the Gray-Scott model. The patterns are obtained on a $128 \times 128$ grid    at $t=15000$ with $D_u=0.2$ and $D_v=0.1$. (b) Initial condition for the simulations. Local perturbations (a convex square and a concave $\Gamma$-shape) are randomly placed, then globally shifted and rotated (assuming periodic boundaries). (c) Illustration of the forward time centered space method. (d) Three consecutive patterns with $\Delta t=30$ are combined into a single three-channel RGB image. These images will appear gray for stationary patterns and  colored for dynamic patterns.}
\end{figure*}

\section{The Gray-Scott Model}
The Gray-Scott Model~\cite{Gray1983autocatalytic, Pearson1993} is a system of two coupled reaction-diffusion equations with two scalar concentration fields $u$ and $v$,
\begin{align}
\label{Eq1}\frac{du}{dt} & = D_u \nabla^2 u - u v^2 + f (1-u)  \\
\label{Eq2}\frac{dv}{dt} & = D_v \nabla^2 v + u v^2 - (f + k) v \, ,
\end{align}
where $D_u$ and $D_v$ are diffusion coefficients and $f$ and $k$ are positive reaction rates. The fields $u$ and $v$ can be interpreted as the concentrations of two reactants that diffuse and react such that they catalyze or inhibit each other's creation.~\cite{Murray2002} To explore the system's behavior, we need to take into account three independent parameters, the ratio of diffusion coefficients (one coefficient can be absorbed into the spatial derivative and changes only the length scale of the pattern) and the two reaction rates $f$ and $k$.

The roots of the reaction equations are the homogeneous steady solutions for which the time and spatial derivatives are zero. Because the reaction equations are third-order polynomials, there are up to three real homogeneous steady states. It is straightforward to show that these correspond to
\begin{equation}
	(u,v) = 
	\begin{cases}
		(1,0)\\
		\label{EqS}\left(\dfrac{\pm \sqrt{C}+f}{2f}, \dfrac{-(\pm\sqrt{C}-f)}{2k+2f} \right)\\
	\end{cases},
\end{equation}
where $C=-4fk^2-8f^2k-4f^3+f^2$.~\cite{Mazin1996}  The trivial solution $(u=1,v=0)$  is independent of $f$ and $k$. The other two solutions exist only if $\sqrt{C}$ is real-valued, which for  $f>0$ and $k>0$ occurs for $k\leq(\sqrt{f}-2f)/2$ and $f\leq1/4$. 

For certain conditions these homogeneous solutions are unstable against perturbations for some parts of  parameter space.~\cite{Murray2002,Murray2003} We can  use linear stability analysis, for which the reaction term is linearized around the steady state, to determine if a perturbation with a certain wavelength grows or decays.~\cite{Murray2002} 
Due to the nonlinear reaction kinetics, these instabilities do not always grow to infinity and the system exhibits more complex behavior, for instance, a variety of stable spatiotemporal patterns occur for certain parts of the parameter space when changes are introduced in the initial conditions.  This behavior was  observed numerically  by Pearson.~\cite{Pearson1993} Twelve heterogenous steady and time-dependent solution classes were identified initially and further investigations revealed more possible classes.~\cite{Munafo2014stable}

Figure~\ref{FigureClasses}(a) shows several typical examples of classes, with  the system being perturbed by a small number of initial ``excitations'' [an example of the initial state is shown in Fig.~\ref{FigureClasses}(b)].  The value of $u$ is displayed in the patterns in Fig.~\ref{FigureClasses}(a). The initial conditions evolve into several different classes of patterns, including fast spatiotemporal dynamics with moving fronts and self-replicating spots, stationary heterogeneous patterns,  and mixed patterns with stationary parts and localized regions of activity. These classes are labeled by Greek letters $\alpha$ to $\nu$, plus two stable homogeneous steady states from Eq.~\eqref{EqS} with low (L) and high (H) concentrations. A complete overview of all classes is given in Sec.~1 of Ref.~\onlinecite{Scholz2021d}. Some classes produce visually similar snapshots, but differ in their time-dependence as demonstrated in video~1 of Ref.~\onlinecite{Scholz2021c}. An online program for exploring the Gray-Scott model in real time is available in Ref.~\onlinecite{Munafo2021}.

We can solve Eqs.~\eqref{Eq1} and \eqref{Eq2} numerically using the forward time centered space method.~\cite{Press2007numerical,Ji2016} The fields are discretized on a $128 \times128$ grid, such that $x=j\Delta$, $y=i\Delta$, $t=n\delta t$ and $i,j,n\in\mathbb{N}$. Differential operators are replaced by forward and central finite differences with spacing $\Delta=1$ (length units) and time step $\delta t = 0.25$ (time units), as illustrated in Fig.~\ref{FigureClasses}(c). The discrete form of Eq.~\eqref{Eq1} [there is an 
analogous equation for Eq.~\eqref{Eq2}] is
\begin{equation}
	u_{i,j}^{n+1} = u_{i,j}^{n} + \frac{D_u  \delta t}{\Delta^2} \left(u_{i+1,j}^{n}+u_{i-1,j}^{n}+u_{i,j+1}^{n}+u_{i,j-1}^{n}-4u_{i,j}^{n}\right)+\delta t\, r(u_{i,j}^{n},u_{i,j}^{n}),
\end{equation}
where $r(u,v)$ denotes the reaction terms. Because each  time step   depends on the previous one, we can integrate the solutions via simple loops over all indices. We assume periodic boundary conditions so that indices are wrapped at the system edges. The performance and numerical accuracy are sufficient for our purposes.~\cite{Press2007numerical} The code is available in Ref.~\onlinecite{Scholz2021} (see \texttt{Gray\_Scott\_2D.py} and \texttt{Generate\_Data.py}).

These numerical solutions are the inputs for the convolutional neural network. To distinguish time-dependent solutions from stationary solutions, we store three consecutive snapshots (with a time-interval $\Delta t$ of 30), as shown in Fig.~\ref{FigureClasses}(d). For each realization of the simulation we store an array of size $128^2\times3$, which can be displayed as a RGB color image. Stationary patterns will appear gray, while time-dependent patterns will show red and blue shifts, depending on the change in concentration, see Fig~\ref{FigureClasses}(d). The classification of such patterns is then equivalent to the classification of the color images. 

\section{Classification of Patterns via Neural Networks}
We initially classify patterns whose existence and lack of sensitivity to the initial conditions are well described in the literature.~\cite{Pearson1993, Mazin1996, Wei2003, Munafo2014stable} Additional pattern types have been identified in Ref.~\onlinecite{Munafo2014stable} for slightly more complicated initial conditions.  The known classes have been mostly identified  visually and their descriptions are often only semantic, making it difficult to determine where certain types of patterns are found in  the parameter space, because a large number of values of the parameters need to be investigated.

We could attempt to define quantities that reduce the patterns and make it easier to identify distinct classes, but such definitions are not simple. Possibilities include Fourier transforms,~\cite{Ouyang1991} statistical methods,~\cite{Wolfram1984universality} and integral-geometric analysis.~\cite{Mecke1996, Schturk2010, Guiu2012, Scholz2015} The drawback of these methods is that they only work in special cases and cannot be generalized easily.

An alternative way to classify images is to use a CNN to fit the entire dataset to pre-labeled data.~\cite{Bishop2006pattern}
Because we already have a classification of patterns with corresponding parameters,~\cite{Pearson1993,Munafo2014stable} we can use this knowledge to train a neural network to classify patterns automatically.
In this section we will introduce the basic terminology. We illustrate how to construct and fit a simple neural network model to data. Then we explain how to design a complex neural network with convolutional layers that is suitable for classifying patterns in the Gray-Scott model.

\subsection{Linear regression}

We first discuss a simple linear regression problem in physics.\cite{Montgomery2017} 
Consider a free fall experiment, where we drop an object and measure its positions $p_0, p_1, p_2, p_3, \ldots$ after times $t_0, t_1, t_2, t_3, \ldots$\,. We know the object is accelerated by gravity, so the expected relation between position and time is $p_i = -\frac{1}{2} g t_i^2 + p_0 +  \epsilon_i$, where $g$ is a fit parameter that we want to determine ($g$ is our estimate of the gravitational acceleration) via fitting the model to the data. The position at $t_0=0$ is $p_0$ and the measurement uncertainty is described by a normally distributed random variable $\epsilon_i$, which is called the error or residue. The model is  linear, because it is linear in the unknown parameter $g$. It can be shown that the most likely estimate for $g$ can be obtained by the method of least squares. Hence, we search for a value of $g$ that minimizes the sum of the squared deviations $\sum_i \left(p_i - \left(p_0-\frac{1}{2} g t_i^2\right)\right)^2$.\footnote{For linear models this is done by computing the minimum norm solution to a set of linear equations. For nonlinear models we typically need to approximate the solution iteratively starting from an initial estimate of the parameters.} After determining $g$ from the fit, we can use the model to predict the position at times that we did not explicitly measure. Here time is a predictor variable and position is  a response variable; the actual time values are the input and the values of the position are the output of the model.\cite{Montgomery2017}

Neural networks are a related concept that can be applied to more general problems. For instance, for predicting patterns in the Gray-Scott model linear regression is not feasible for several reasons. First, there are too many predictor variables, in our case $128^2\times3$, which would require too many free parameters.\footnote{For $128^2\times 3 = 49152$ predictor variables, even if we  include only zeroth and first order terms ($2\times 49152$) and two factor interactions (all possible products of two different predictor variables $0.5\times 49152 \times 49151$), we would need to fit more than $1.2$ billion parameters.} Second, we have a classification problem, where the responses are classes and not numbers. Third, we do not know the relation between predictors and the responses because the relation is very complex. However, we can generate many realizations of patterns from known classes and use it to train (i.e., fit) a model. 

\subsection{Fully connected neural networks}

In the following, we discuss the basics of convolutional neural networks and how they can be applied to classify patterns in the Gray-Scott model. Neural networks are models of nested (linear, logistic, or nonlinear) regression equations that mimic networks of biological neurons.~\cite{Hopfield2554} 
Each neuron is a continuous real variable and neurons are structured into several layers that are connected via a (nonlinear) transformation. For classification problems, the purpose of the layers is to gradually map a high-dimensional input (the many predictor variables) to a low-dimensional output of one neuron per class, each of which returns the probability that the input belongs to that class. A set of predictor variables $x_1^{(0)}\dots x_D^{(0)}$ forms the input layer of the neural network. In contrast to our linear regression example, the subscript index now refers to different neurons and not to individual measurements or samples. A layer of a neural network refers to all $x_j^{(n)}$ values with equal $n$. All neurons, i.e., values of a deep layer $(n+1)$ are calculated from the values of the previous layer $(n)$ by the transformation:
\begin{equation}
\label{EqNN} x_j^{(n+1)} = h\left(\sum_{i=1}^{D} w_{ji}^{(n+1)} x_i^{(n)} +w_{j0}^{(n+1)} x_0^{(n)}\right),
\end{equation}
where $w_{ji}^{(n+1)}$ are the free parameters of the model, known as weights. 
By  convention, the superscript for weights starts at $1$, which hence connects the input layer to the first layer.
The quantity $w_{j0}^{(n+1)} x_0^{(n)}$ is a bias term required for nonzero intercepts, similar to the constant term required for a polynomial that does not intercept the origin. 
The sum in Eq.~\eqref{EqNN} is simply a linear combination of the input neurons, called the activation $z_j^{(n)}$. This quantity is the argument of the activation function $h$,\cite{Bishop2006pattern} which generates the values of the neurons in layer $(n+1)$.

\begin{figure}[tb]
	\includegraphics[width=0.8\columnwidth]{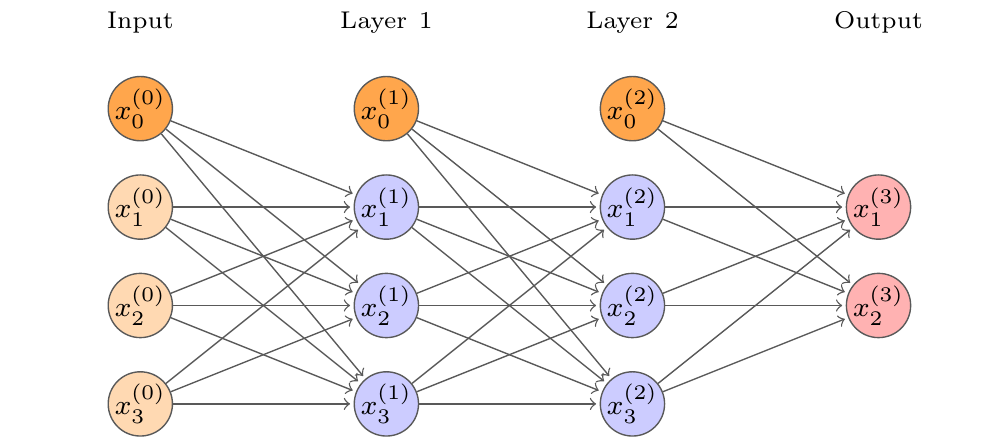}
	\caption{\label{nnet} Illustration of a simple fully connected neural network with three input values, two hidden layers with three neurons per layer and two output classes. Each layer contains a bias unit to allow for nonzero intercepts. Each arrow corresponds to a trainable weight. The hidden layers allow the network to be trained, but because these layers typically do not represent a simple relation, they are not reported as output.  Note that   every neuron is connected to every other neuron in the previous layer.}
\end{figure}

The choice of activation functions depends on the type of problem.~\cite{Ramachandran2017searching} For deep layers we use the rectified linear unit ({\tt ReLU}) function $h(z_j^{(n)})=\max(0,z_j^{(n)})$, which ensures a neuron can be in an inactive ($0$) or active state ($>0$). For hidden layers this quantity is difficult to interpret. Simply put, each neuron in a layer can be activated by certain linear relations in the previous layer, but in particular for deep layers $(n\geq1)$ it is not known how to generally interpret these values. The outputs of these layers are typically not reported and therefore are called hidden.

The last layer of the neural network returns the response. For classification problems, the number of neurons in this layer must be equal to the number of classes and each neuron should return a probability that a pattern belongs to a certain class. To output these probabilities we use the {\tt softmax} activation, which is the normalized exponential function $h(z_j^{(n)})=\exp(z_j^{(n)})/\sum_j{\exp(z_j^{(n)})}$. We interpret the output as probabilities because by definition it is between zero and one and all values add up to one. Because every neuron is connected to every other neuron in the previous layer in simple neural networks, they are called dense or fully connected layers. Equation~\eqref{EqNN} defines the model. The weights $w_{ij}^{(n+1)}$ are the fit parameters and are determined by fitting the model to the training data, which in our case are the individual patterns and their corresponding classes. Training a neural network means adjusting the weights by fitting the model to the training data to optimize the prediction accuracy. The weights are set randomly at the start of the training and are then iteratively updated during optimization.

The goal of the optimization is to minimize the loss function, which quantifies how close the predicted responses are to the correct class. For classification problems, the responses are probabilities and the loss function is not the sum of square deviations, but the sparse categorical cross-entropy defined as
\begin{equation}
\label{Eq_CE}
\mbox{CE} = -\sum\limits_{l=1}^{L}\sum\limits_{k=1}^{K} \tau_{lk} \log\left(p_k(\mathbf{\bar{x}}_l)\right),
\end{equation}
where $L$ is the number of input samples $\mathbf{\bar{x}}_l$ and $K$ is the number of classes. The ($L\times K$) sparse matrix $\tau_{lk}$  contains a single entry per row  corresponding to the correct class of each sample. The probabilities $p_k(\mathbf{\bar{x}}_l)$ are the output of the last layer (after  softmax activation) and depend on the weights. We call these the predictions that a sample $\mathbf{\bar{x}}_l$ is of class $k$. By definition, $\sum_{k}p_k(\mathbf{\bar{x}}_l) = 1$, so that the predictions can be identified as probabilities. A perfect prediction would mean $p_k(\mathbf{\bar{x}}_l)=1$  if $k=1$   for the correct class, and therefore $\mbox{CE}=0$. Hence, if $p_k(\mathbf{\bar{x}}_l)<1$, we have $\mbox{CE} > 0$. A complete discussion of the optimization procedure is beyond the scope of this article. We will give an explanation of the method in Sec.~\ref{sec:previous3}.

Let us consider a fictitious example. A group of mathematics and physics students take an exam with three problems P1, P2, and P3. Each problem is graded with up to 10 points. Some outcomes of the exam are illustrated in Table~\ref{tab:table2}. We want to use a neural network to distinguish mathematics and physics students from their scores (for    simplicity we assume these groups are mutually exclusive). We train a neural network with two hidden layers, each with three neurons, as shown in Fig.~\ref{nnet}. There are three predictor variables and two classes, i.e.,  two output neurons for the response. The training data consists of individual scores and the class of each student. The responses of our model will be probabilities that a student is a mathematics or a physics student.

\begin{table}[b]
	\caption{\label{tab:table2}
Fictitious example for a simple classification problem. Given exam scores (three problems P1, P2, P3) of mathematics (M) and physics (P) students, we assume a neural network model that predicts the type of student from the exam scores. The values are just illustrative and do not correspond to a real world example. In a trained model, predictions should be as close as possible to the actual classes. Here, for student 4 the prediction does not fit the true class, indicating a problem of the model or of the dataset.}
		\begin{tabular}{*{6}{|C{.12\textwidth}}|*{2}{C{.12\textwidth}}}
		\cline{1-6}
		  \multicolumn{6}{|c|}{Training data}& &\\
		 \hline
		 \multicolumn{1}{|c}{Student}&
		 \multicolumn{3}{|c|}{Predictor (scores)}&\multicolumn{2}{c|}{Response (classes)}&\multicolumn{2}{c|}{Model prediction after fit}\\
		\hline
		No.&P1&P2&P3&M&P&\multicolumn{1}{c|}{M}&\multicolumn{1}{c|}{P}\\
		\hline
		1&9&8&10&1&0&\multicolumn{1}{c|}{0.8}&\multicolumn{1}{c|}{0.2}\\
		\hline
		2&1&5&10&0&1&\multicolumn{1}{c|}{0.3}&\multicolumn{1}{c|}{0.7}\\
		\hline
		3&3&7&2&1&0&\multicolumn{1}{c|}{0.6}&\multicolumn{1}{c|}{0.4}\\
		\hline
		4&3&7&3&0&1&\multicolumn{1}{c|}{0.55}&\multicolumn{1}{c|}{0.45}\\
		\hline
		$\vdots$&$\vdots$&$\vdots$&$\vdots$&$\vdots$&$\vdots$&\multicolumn{1}{c|}{$\vdots$}&\multicolumn{1}{c|}{$\vdots$}\\
		
		\end{tabular}
	
\end{table}

We observed that for most students the predicted probabilities are in accordance with the true class, that is, the probability is largest for the true class. However, we also see exceptions, indicating that the model was not able to fully fit the training data. For instance, students three and four achieved similar scores, but belong to different classes. This could mean we do not have enough data to reliably distinguish such subtle differences or even that the predictor variables are not reliable to distinguish mathematics and physics students.

\subsection{Convolutional neural networks\label{sec:previous}}

Although fully connected neural networks can in principle mimic any data operation, it is difficult to train such a model to perform meaningful operations for datasets with large dimensions and complex correlations, such as images or audio signals. This problem is called the curse of dimensionality.~\cite{Bishop2006pattern} Our patterns consist of many predictor variables, namely all the input concentration values, here $128^2\times3$. However, we know, by definition in two-dimensional patterns at least, nearby values are correlated. But the correlations can be very complicated. One way to design a neural network that can learn such complicated correlations is to introduce convolutional layers. For these layers, the neurons are replaced by convolutions of the input data with convolution kernels. The values in the convolution kernel are the free parameters that are optimized in the training process, similar to the weights in the fully connected network.

\begin{figure}[tb]
	\includegraphics[width=\columnwidth]{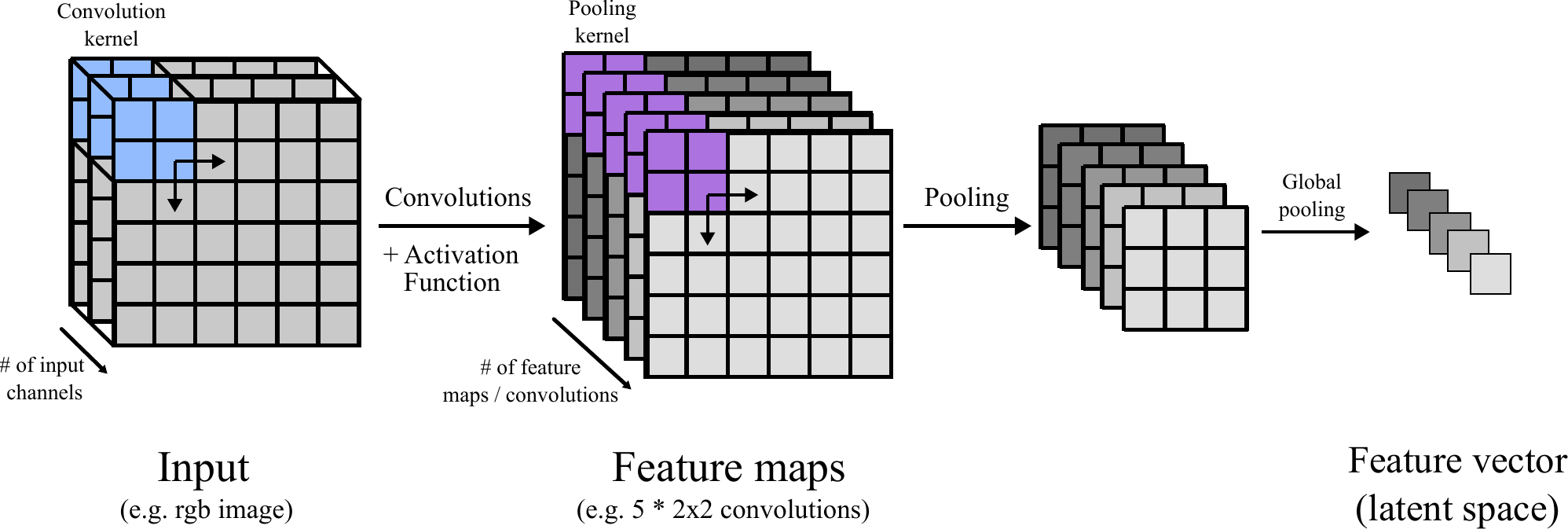}
	\caption{\label{ConvPool} Illustration of a two-dimensional convolution and pooling operation with a three-channel input image. The convolution kernel size needs to be defined in the horizontal and vertical directions, but the depth always corresponds to the number of channels in the input. Here we use five convolution kernels of size $2\times2\,(\times3)$, which results in 5 feature maps. To reduce the size we apply maximum pooling with a $2\times2$ kernel. This kernel chooses only the maximum value in successive $2\times2$ patches, and therefore reduces the total size of each map by 1/4. In a last step, we perform a global maximum pooling, which picks only the maximum value of each feature map, yielding a five-dimensional feature vector. Such a vector can then be used as input to a dense neural network for  classification.
	}
\end{figure}

The number and size ($ \mbox{width}\times \mbox{height}$) of the convolution kernels are design parameters of the network. Each convolution kernel is shifted by the stride $s$ across the input image, calculating the product of the convolution with sub-part of the image. The result is a feature map. In Fig.~\ref{ConvPool} we show an example with five convolution kernels, that is, five resulting feature maps. Although the size of the kernel is a design parameter, the depth of each kernel is always equal to the number of channels in the previous layer. For the input layer these channels are often the RGB values of an image, but in our case they are simply the three time steps. Such a layer is called a two-dimensional convolution layer. At the boundary of the system we can either apply padding, for instance, fill in zeros, to keep the output size the same as the input, or discard convolutions at the boundary, slightly reducing the output size depending on the kernel. If the size of the input patterns is large and the resolution is high enough, we can increase the stride $s$ by which the convolution kernel is shifted across the system to reduce the computational cost. As for fully connected networks, an activation function is applied to the feature maps.

By using the output of convolutional layers we can reduce the dimensionality of the problem by extracting only relevant features from the images and then feeding them as inputs to deeper layers of the network.  To do so, a coarse-graining operation called pooling is applied to the feature maps. In this operation, a pooling kernel is shifted across the map (with a stride equal to the dimensions of the kernel) and selects only the maximum inside its range in the case of max pooling. For instance, a pooling kernel of size $2\times2$ reduces the size of the map by a factor of four. A simple example of feature extraction via a convolution and max pooling operation for the pattern from Fig.~\ref{FigureClasses}d) is shown in Sec.~2 of Ref.~\onlinecite{Scholz2021d}. To convert all feature maps into a single feature vector, a global max pooling is typically used, where from each feature map only the maximum value is selected. These feature values must be separable into linearly independent components, such that neurons in the last layer can separate all classes. 

To reduce the impact of overfitting, dropout layers can be used during training. These operations randomly set a certain fraction of weights to zero during the training which prevents single neurons from having a large impact on the classification result.

Due to the popularity of neural networks, there are  well-maintained implementations of neural networks available. These libraries make it easy to design a neural network architecture without the need to implement the underlying low-level algorithms. We use Python and the Keras library.~\cite{chollet2015keras} The corresponding python code for generating the input data as well as the training and verification is available.~\cite{Scholz2021} In the next section, we  describe the architecture of the corresponding neural network independently of its implementation. All steps are described at a high level, so that our example can be applied to other neural network implementations, such as Pytorch~\cite{pytorch} and the Matlab deep learning toolbox.~\cite{MatlabDLT}

\subsection{Deep neural network architecture}

In   Sec.~\ref{sec:previous} we described the fundamental layers required for a convolutional neural network. In practice, many of these operations have to be executed successively, which is why such networks are called deep convolutional neural networks. The architecture of our neural network is listed in Table~\ref{tab:table1}.

\begin{table}[tb]
	\caption{\label{tab:table1}
		Architecture of the two-dimensional convolutional network with layer type, size and stride $s$ of the convolution kernel, activation function, and the number of parameters (fixed and trainable). The stride $s$ is the amount of pixel shift of the convolution kernel. Conv2D refers to a two-dimensional convolution, MaxPool is a maximum pooling operation and GlobalMaxPool refers to the global maximum pooling. Dense layers are equivalent to a fully connected neural network. The dropout layer is  active only during training and randomly sets $15\%$ of the weights to zero in each iteration. For the last layer a Softmax activation is applied to convert the output into probabilities. }
	\begin{ruledtabular}
		\begin{tabular}{lccrr}
			\textrm{Layer}&
			\textrm{Kernel size}&
			\textrm{Output}&
			\textrm{Activation}&
			\textrm{Parameters}\\
			\colrule
			Batch normalization & -- & 128, 128, 3 & -- & 12\\
			Conv2D & 3,3\,(2s) & 63, 63, 32 & ReLU & 896\\
			Conv2D & 3,3 & 61, 61, 64 & ReLU & 18496\\
			MaxPool & 2,2 & 30, 30, 64 & -- & 0\\
			Conv2D & 3,3 & 28, 28, 128 & ReLU & 73856\\
			MaxPool & 2,2 & 14, 14, 128 & -- & 0\\
			Conv2D & 3,3 & 12, 12, 128 & ReLU & 147584\\
			MaxPool & 2,2 & 6, 6, 128 & -- & 0\\
			GlobMaxPool & -- & 128 & -- & 0\\
			Dense & -- & 256 & ReLU & 33024\\
			Dropout & -- & 256\,(85\%) & -- & 0\\
			Dense & -- & 15 & Softmax & 3855\\
			\colrule
			& & & & $\sum$=\,277723\\
		\end{tabular}
	\end{ruledtabular}
\end{table}

We use a series of two-dimensional convolutions and pooling operations, as described in Fig.~\ref{ConvPool}. The number of parameters (weights) per convolution layer is given by [W$\times$H of the convolution kernel] $\times$ [the number of convolution kernels] $\times$ [the depth of the previous layer] $+$ [the number of bias units\,(equal to the number of convolution kernels)]. The input to the network is normalized per batch, such that the prediction is independent of the scale of the input. A batch here is a subset of the training data that is consisting of the input images and the according classes. One batch at a time is processed through the CNN during a single optimization step, until all batches are processed. The reasons for not using the full training set at once are memory constraints, i.e., not being able to load the whole dataset into memory, and the increase in training speed as the weights are updated after each batch. However, if the batch size is too small, the batch might not be representative of the entire data set. We use a batch size of 128 data sets in our code.

To define this neural network in Keras, the following python code is used
\begin{verbatim}
	model = tf.keras.Sequential()
	model.add(tf.keras.Input(shape=shape_input))
	model.add(tf.keras.layers.BatchNormalization())
	model.add(tf.keras.layers.Conv2D(32, (3, 3), strides=2, activation="relu"))
	model.add(tf.keras.layers.Conv2D(64, (3, 3), activation="relu"))
	model.add(tf.keras.layers.MaxPooling2D(2, 2))
	model.add(tf.keras.layers.Conv2D(128, (3, 3), activation="relu"))
	model.add(tf.keras.layers.MaxPooling2D(2, 2))
	model.add(tf.keras.layers.Conv2D(128, (3, 3), activation="relu"))
	model.add(tf.keras.layers.MaxPooling2D(2, 2))
	model.add(tf.keras.layers.GlobalMaxPooling2D())
	model.add(tf.keras.layers.Dense(256, activation="relu"))
	model.add(tf.keras.layers.Dropout(0.15))
	model.add(tf.keras.layers.Dense(num_categories))
\end{verbatim}
A minimal example with the full code is discussed in Sec.~3 of Ref.~\onlinecite{Scholz2021d}. First, a model of class sequential is created. Then all necessary layers are added using the \texttt{model.add} function. Four layers act as pooling layers (MaxPooling2D) to gradually reduce the dimension of the data after convolutions. As illustrated in Fig.~\ref{ConvPool} the max-pooling layer keeps only the maximum within a pooling kernel, and a global max-pooling layer finally propagates only the maximum of each convolution feature to the following dense layer.   Each feature is then represented by a single value and the dense layer only needs to distinguish the output classes in terms of activations from the pooled final convolutional feature layer. This layer has an output size of  128, compared to the $3\times 128^2$ input variables, which mitigates the curse of dimensionality and  makes it  easier to optimize the final dense layers to properly distinguish different classes. Note that  for the last layer we do not need to explicitly set the activation to softmax, because it will be applied during the optimization step automatically.

\begin{figure*}[tb]
	\includegraphics[width=\textwidth]{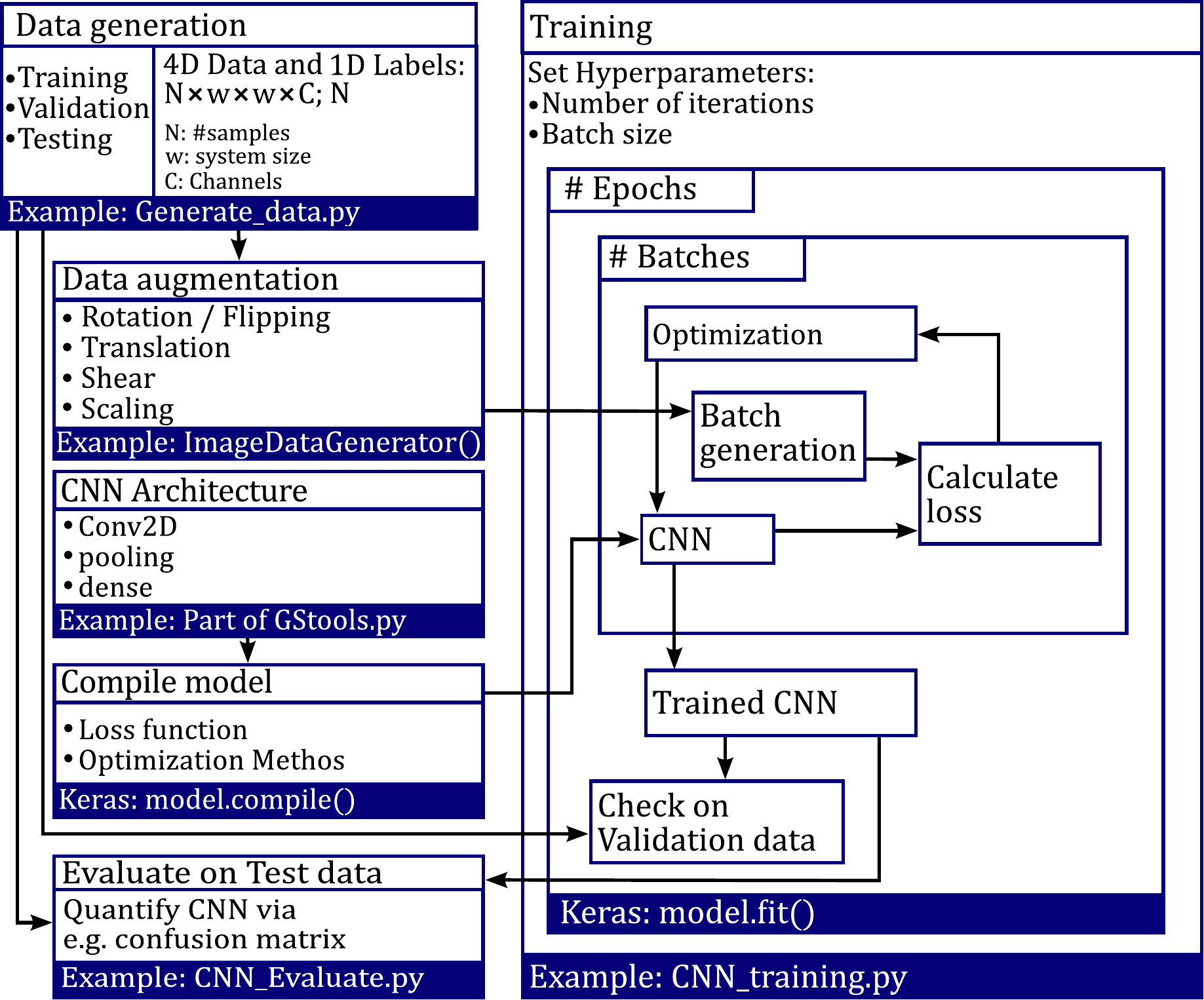}
	\caption{\label{FigureFlow} Illustration of the data generation, training, validation, and test procedure. This workflow is common for high level machine learning. The generated datasets serve as input to the data augmentation function, validation, and test step. The CNN is defined and the weights of the network are iteratively optimized in the training loop by minimizing the categorical cross-entropy. After the training is finished, the predictive capabilities of the CNN are evaluated with the test data. The code examples correspond to the supplementary material in Ref.~\onlinecite{Scholz2021} and, if applicable, the most important Keras function in each box. For a more detailed description see  the Appendix. }
\end{figure*}

In Fig.~\ref{FigureFlow}  we summarize the  procedure of training and validating the neural network and show a flow chart of the  algorithm with references to the our code.~\cite{Scholz2021}

\subsubsection{\label{sec:previous2} Generate training, validation and test dataset} 

It is convenient  to   split the available data into training, validation and test data. The training dataset is used for the actual optimization. Predictions of the independent validation dataset are used during the training to check if the performance generalizes beyond the training data. If the accuracy is high for the training data, but low for the validation data, this is typically a sign of overfitting. Overfitting can be prevented by manual optimization of the architecture of the neural network, for instance, by reducing the number of parameters that are trained in the neural network. To make sure no overfitting occurs, another independent test dataset is used to evaluate the accuracy of the network after the training.

As we have described, we vary the initial conditions for each pattern and solve the Gray-Scott model numerically to obtain three snapshots of the system at $t=14940, 14970$ and $15000$. 
These  values were chosen because the system has converged to a steady state based on visual inspection, see also Ref.~\onlinecite{Scholz2021c}. We generate a total of 1500 samples, 100 per class (each with a unique pair of the reaction rates $f$ and $k$), split evenly between the training and validation datasets. For the test dataset we generate another set of 750 patterns, but  slightly vary  $f$ and $k$ to test how well our predictions generalize to other parts of the parameter space. Each dataset is stored in a four-dimensional array (see the Data generation box in Fig.~\ref{FigureFlow}). Because data generation can require a  long time, we also provide a pre-generated set of data.\cite{Scholz2021b}

\subsubsection{\label{sec:previous3}Training and optimization}
With the test and validation data sets we iteratively optimize the neural network weights by minimizing the categorical cross-entropy in Eq.~\ref{Eq_CE}. One iteration over the entire training dataset is called an epoch. 
The training data contains 750 patterns, but the model has many more free parameters (see Table~\ref{tab:table1}). To reduce the danger of overfitting, the training data is parsed by a data augmentation function, which selects batches from the training data during each epoch and applies random rotations, translations, shearing and scaling to the input to artificially augment the available training data. The categorical cross-entropy is minimized during this process using a stochastic gradient descent method. The gradient of the categorical cross-entry with respect to the weights can be calculated efficiently using  back-propagation,~\cite{Bishop2006pattern} which is  implemented in Keras by the TensorFlow backend.~\cite{tensorflow2015-whitepaper} After each epoch, the performance of the neural network is compared to the validation dataset and recorded for later evaluation. For an example see also Sec.~3 of Ref.~\onlinecite{Scholz2021d}.

\subsubsection{Evaluation of accuracy} 
\begin{figure*}[tb]
	\includegraphics[width=\textwidth]{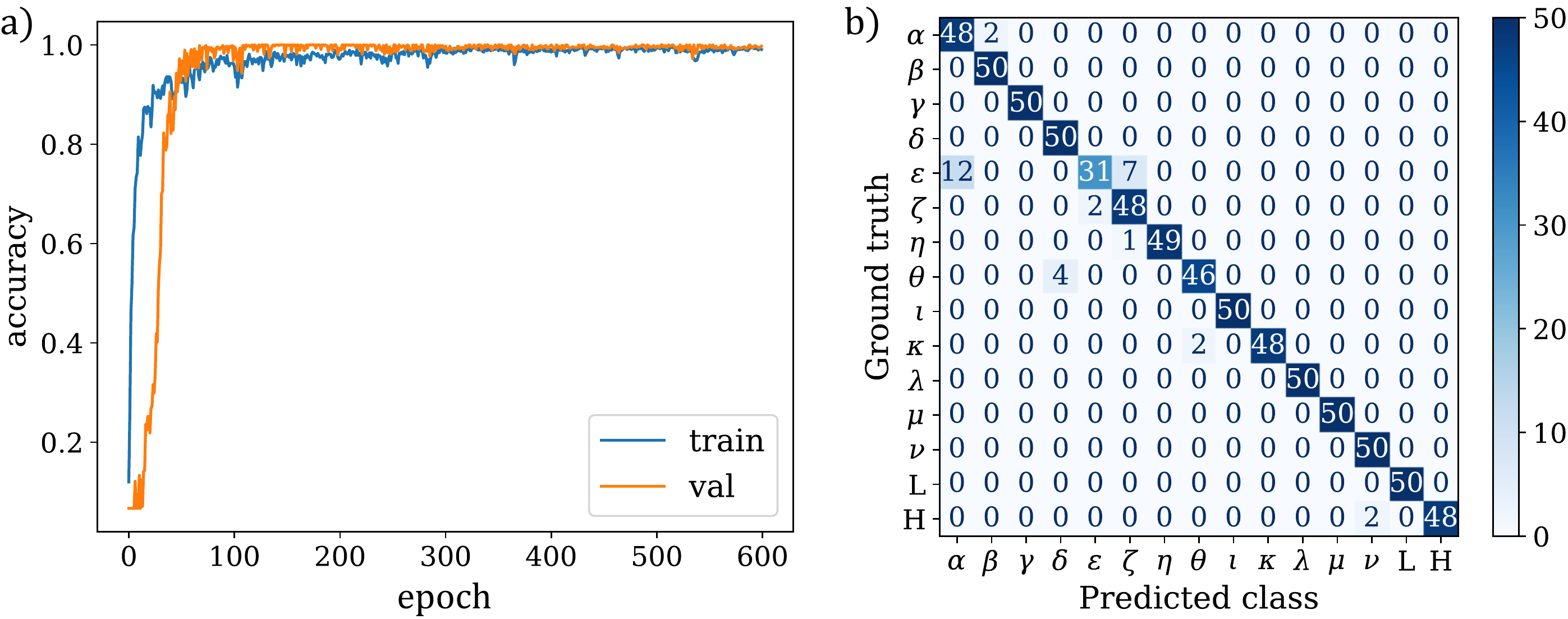}
	\caption{\label{FigureTraining} (a) Training history and accuracy for the training and validation data. The accuracy usually approaches    $>99\%$, and we stop the training phase after 600 epochs. (b) Confusion matrix for the training dataset; $\approx 96\%$  of the   patterns are correctly classified.}
\end{figure*}

Training and validation accuracy is displayed during training (Training box in Fig.~\ref{FigureFlow}) and is  monitored initially. If the performance is not adequate, meaning it does not converge for both the training and validation set at high accuracy, the architecture of the neural network can be adjusted and the accuracy of the new network is again tested  using  the training and validation data set.\footnote{There are also some constants that affect the optimization procedure. For instance the learning rate, i.e., the step size of the gradient descent. Such parameters are called hyperparameters. Often these hyperparameters are automatically determined or can be kept at default values, but sometimes adjustments might be necessary. For instance, the network might not have enough feature maps, neurons and layers to fit the data. A frequent problem is also low quality of training data due to human error, which requires manual revision. For example, patterns might have been mislabeled. Or the training data might include patterns that do not represent a class very well.} In Fig.~\ref{FigureTraining}(a) we show the training history of the optimization process. For the training and validation dataset we  approached an accuracy of more than 99\%.

We can now check the predictive power of the network on the test dataset for a final evaluation of the accuracy. A simple script to determine predictions of the trained model is shown in the appendix. We can evaluate the classifications of the test dataset in a confusion matrix, which plots the correct classes versus the predicted classes by the model (see Fig.~\ref{FigureTraining}(b)). The majority of the patterns are correctly classified with an  accuracy of about 96\%. There is some misclassification for classes $\epsilon$ as $\zeta$ and $\alpha$. This mismatch is reasonable, because these classes share some similarities, and there could be a gradual transition from one class to the other. One might  say that the neural network disagrees in some cases with the human curator of the dataset.

It is not known how to obtain an intuitive understanding of how the neural network actually performs a task. For this reason neural networks are sometimes also refereed to as black box models. It can be useful to display the saliency maps,~\cite{Kadir2001saliency, simonyan2013deep} which display areas of the input images that have a particularly strong influence on the gradient of the loss function. Examples and a demonstration are included in Sec.~4 of Ref.~\onlinecite{Scholz2021d} and in Ref.~\onlinecite{Scholz2021}.

\section{Results}

We now demonstrate the results of the neural network for predicting classes and describe how we can use the initial results to improve the training data for a second training iteration.

\subsection{Scan of the parameter space}

Our training data  includes only 15 unique pairs of $f$ and $k$ parameter values given for each class.~\cite{Pearson1993,Munafo2014stable}  We now want to use our CNN to scan a much larger and denser parameter space of $f\in[0.02, 0.08]$ and $k\in[0, 0.12]$. This part of the parameter space is selected based on the existence criterion for stable solutions in Eq.~\eqref{EqS}.
Each pattern from this large parameter space represents a different pair $(f,k)$. A total of 5481 patterns are created and stored in a four-dimensional array, where the first dimension corresponds to different pairs of parameters and the others are the three stacked solutions at three consecutive time steps. Obviously, here we do not know the labels a priori and none of the parameter values were explicitly included in the training data. Instead we want to predict the classes by our CNN.

\begin{figure*}[tb]
	\includegraphics[width=\textwidth]{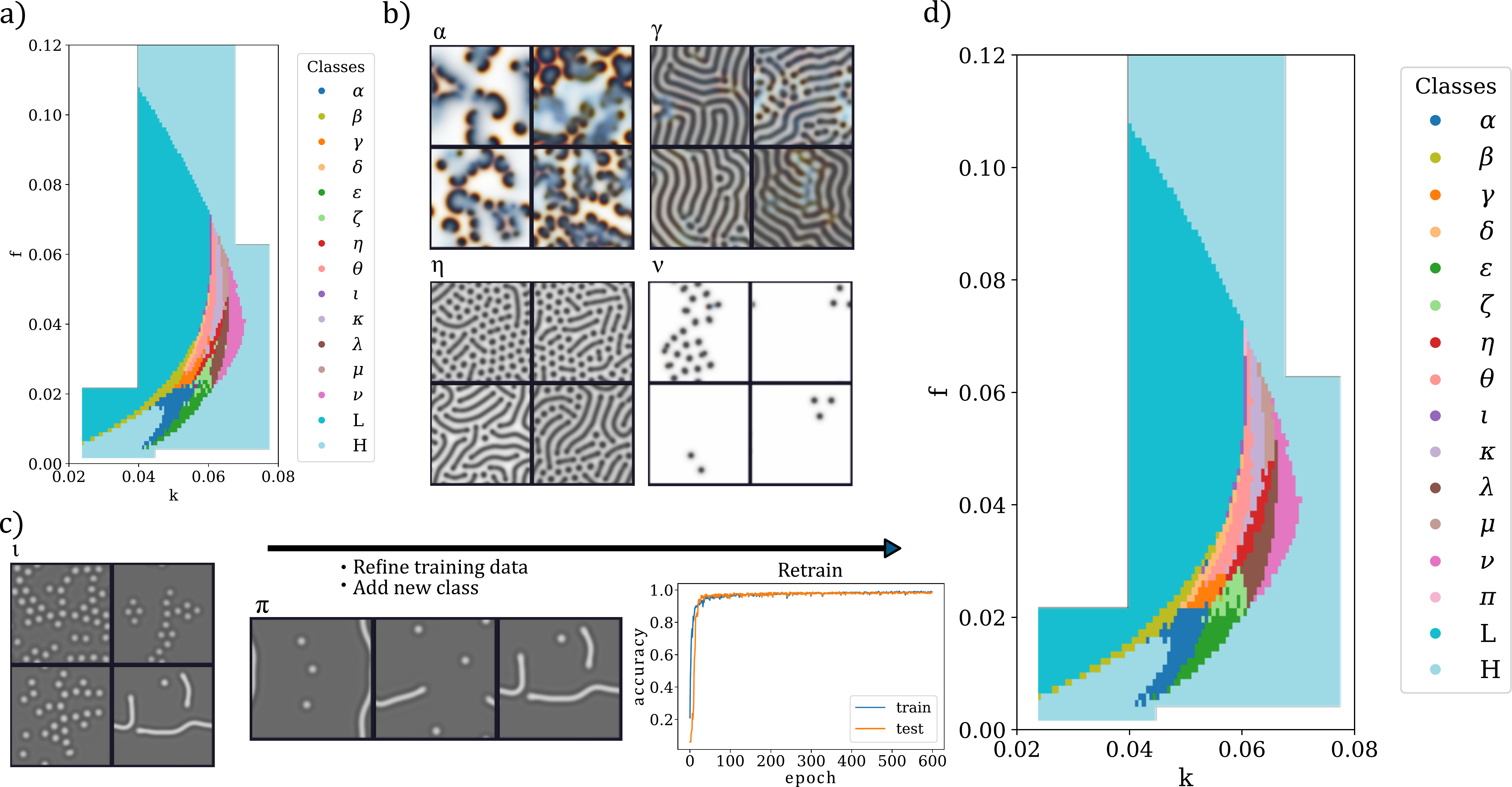}
	\caption{\label{FigureRetrain} (Color online)  (a) Parameter space scan and results of the CNN classification: Connected regions of similar classified patterns are typically observed. Examples for visual inspection of representative patterns from the classification results for the $\alpha$, $\gamma$, $\eta$, $\nu$ and $\iota$ classes. Most patterns are correctly classified, but some clear outliers are observed. We use the information of these outliers as input to improve our training dataset for a second training pass. For instance, some sparse hole patterns are incorrectly labeled as $\nu$, although they behave more like very slowly evolving $\lambda$. (b) A small region of the parameter phase forms a new class $\pi$ that was not taken into account in the first pass of the training. Selecting patterns that fit the definition best lets us create a more refined training dataset to retrain the CNN and perform a better classification. (c) Parameter space scan after a second training pass.}
\end{figure*}

The data is parsed by the CNN using the \texttt{model.predict()} function, which returns the probabilities that a pattern belongs to a certain class. An example for prediction probabilities for a single patterns is shown in the appendix. We select the class with the highest probability for each pattern and plot the results in Fig.~\ref{FigureRetrain}(a). For an example, see \texttt{Parameter\_Space\_Dataset\_Generate.py} and \texttt{Parameter\_Space\_Dataset\_Classify.py} in Ref.~\onlinecite{Scholz2021}. We  see distinct areas for each class, which are  reasonable  compared to the literature~\cite{Pearson1993} (see also Sec.~5 of Ref.~\onlinecite{Scholz2021d}). But how can we truly assess whether the classifications are correct and meaningful? This is the biggest weakness of neural networks, because we have no intuitive understanding of how the neural network performs the classification.

In contrast to manually searching through the entire dataset, we can now pre-sort our patterns by the predicted class and perform a visual inspection of each class. Illustrative examples are shown in Fig.~\ref{FigureRetrain}(a). Most of the patterns are correctly classified. However, in some cases there are outliers that do not match the prediction. It appears that there are mixtures of different classes for which the correct classification is debatable, or even entirely new categories. For instance, as shown in Fig.~\ref{FigureRetrain}(b), we see patterns in the $\nu$ class  with many more dots compared to the training data, but still large homogeneous areas. Upon closer inspection we find that these are very slowly converging $\lambda$ patterns. In the $\iota$ class, we observe some mixtures of dots and individual stripes, as shown in Fig.~\ref{FigureRetrain}(c). These patterns are reminiscent of observations in Ref.~\onlinecite{Munafo2014stable}, where this class was termed class $\pi$. This issue indicates some weakness in the training data. Training data issues can not be solved by optimizing the CNN. Instead we need to improve the training data, as discussed in the next section.

\subsection{Refinement of training dataset and second pass}

Despite    generating proper training, validation and test data, the results of many neural network and similar machine learning models can suffer from poor reproducability when applied to data that was not previously considered in any of the datasets.~\cite{haibe2020transparency} A major problem is spurious correlations in the training, validation, and test data. A well-known example is the presence of watermarks and source labels in image databases, where it was  found that machine learning models can accidentally detect such artifacts and use them for classification instead of the actual objects.\cite{Lapuschkin2019} It is therefore important to assess if predictions made by neural networks are reasonable and robust. In the following  we demonstrate how our prior predictions  can be used to refine the training dataset and make the predictions more accurate.

To improve the training data, we select additional patterns from our previous predictions if a pattern fits the respective class descriptions well and add the corresponding parameters ${f,k}$ to the training dataset generation script. The script \texttt{Predicted\_Class\_Montage.py} in Ref.~\onlinecite{Scholz2021} demonstrates how to interactively display patterns for each predicted class. The corresponding ${f,k}$ parameters are displayed by clicking on the patterns.

We restart the training of our neural network from scratch. Additionally, we add a new class called $\pi$ to the training data to take the misclassified $\iota$ patterns into account. We generate a total of 3200 patterns, 200 for each class including new parameters in some cases, split evenly across the training and validation datasets. The training dataset now contains more pairs $(f,k)$ and more patterns per class. The training accuracy again approaches 99\% [see Fig.~\ref{FigureRetrain}(c)], but displays a better accuracy for the test data. We see faster convergence and slightly fewer fluctuations across the epochs, which confirms the increase in quality of the training data. In Fig.~\ref{FigureRetrain}(d) we show the results of the new classification after the second pass of the iterative training approach. Some quantitative differences are found for classes $\epsilon$, $\zeta$, and $\lambda$, which as we discussed, are  difficult to clearly distinguish. Domains of these classes appear more compact now. The new training data does not fundamentally impact the location of other classes in  parameter space. The new class $\pi$ is robustly detected at $f\approx 0.068$ and $k\approx0.060625$ in agreement with Ref.~\onlinecite{Munafo2014stable}. As expected, these patterns are found in a  small region that is easy to miss without a dense parameter search. Although CNNs cannot directly identify new classes, obvious misclassifications were visible after the first training iteration. These misclassifications are no longer present after the second iteration. Some patterns, in particular those classified as $\beta$ remain difficult to associate uniquely with classes in the current scheme. This difficulty cannot be mitigated by extending the training data, and it might be necessary to add additional classes. Overall, we do not see qualitative changes in predictions after the second training, which is  not   a surprise, because the Gray-Scott model is already well described in the literature. However,   for new systems, we recommend  being very critical of initial results.

\section{Discussion}

We have demonstrated how a CNN can be implemented and used for exploring the patterns in the Gray-Scott model. CNNs circumvent the manual search for characteristic features needed for pattern classification and aid in identifying the boundaries between classes in the parameter space and thus supplement analytical work on reaction-diffusion models. We expect a similar applicability to other reaction-diffusion systems. However, this approach comes at the cost of a lack of intuitive understanding of how patterns are classified by the neural network.

A possible continuation of our approach is to test other types of initial conditions, for which even more classes of patterns can be found.~\cite{Munafo2014stable} One way to achieve more robust classifications is to use pre-trained neural network architectures and adapt them to this use. Such networks are available for pattern recognition trained on large datasets using high performance computing clusters. The feature extractors of such networks can be transferred to new uses by retraining the weights of the last fully connected layer on a new dataset. We also could make more use of the time-dependence in the simulations. Here, we have only demonstrated two-dimensional convolutional layers. However, if we add more channels and therefore more information on the time-dependence to the data, we can use three-dimensional convolutions or sequence classification in the network architecture. An example is discussed in Sec.~6 of Ref.~\onlinecite{Scholz2021d}.

Neural networks can also fit continuous response variables, not just discrete classes. For instance, CNNs can predict solutions to partial differential equations,~\cite{Han8505, Raissi2019physics} and even simple reaction-diffusion equations.~\cite{Li2020reaction} Their versatility makes neural networks very attractive to use for complex data analysis and we expect many more applications in physics. 

\begin{acknowledgments}
We would like to acknowledge inspiring discussions on topological transitions in reaction-diffusion systems with Klaus Mecke and Gerd Schröder-Turk.
\end{acknowledgments}

\appendix*

\section{Explanations of    example scripts}

\noindent To run a single simulation of the Gray-Scott model for a specific set of input parameters [seed] [$D_u$] [$f$] [$k$] execute:
\begin{verbatim}
python Gray_Scott_2D.py 2 0.2 0.009 0.045
\end{verbatim}
The first two parameters are optional. Varying the seed will change the initial conditions randomly. See Sec.~1 of Ref.~\onlinecite{Scholz2021d} or the function get\_dataset\_parameter() in {\tt GStools.py} from Ref.~\onlinecite{Scholz2021} for example parameters. 

To generate training, validation, and test datasets for the 2D convolution CNN, execute the following commands. Depending on the CPU, you can adapt the constant NTHREADS (default 8) to speed up parallel execution.
\begin{verbatim}
	python Dataset_Generate.py train 2D
	python Dataset_Generate.py val 2D
	python Dataset_Generate.py test 2D
\end{verbatim}
Each script requires about two hours on an Intel i7-9750H CPU (2.6\,GHz). These scripts will create the data files \texttt{Dataset\_2D\_train.p}, \texttt{Dataset\_2D\_val.p} and \texttt{Dataset\_2D\_test.p}. All data files   can be  downloaded from an Open Science Framework (OSF) repository at Ref.~\onlinecite{Scholz2021b}.

We can use  the datasets generated by the previous scripts, and then run the following script to train the model:
\begin{verbatim}
	python CNN_Train_2D.py
\end{verbatim}
The training runs for about 1.4 hours. For demonstration purpose the number of epochs can be reduced. The model will be saved in the subfolder  \texttt{model\_CNN\_2D}. Pre-trained models are also stored in Ref.~\onlinecite{Scholz2021}. 

\noindent A simple script to load the test data, the neural network model, and calculate predictions is the following:
\begin{verbatim}
# Import dependencies
import tensorflow as tf
import numpy as np
import pickle

# Load model and dataset
model = tf.keras.models.load_model("model_CNN_2D")
dataset_test = np.array(pickle.load(open("Dataset_2D_test.p", "rb")))

# Calculate predictions of neural network for training dataset
labels_test_pred = model.predict(dataset_test)

# Display predictions for pattern np. 26 in training dataset as probabilities
layer = tf.keras.layers.Softmax()
layer(labels_test_pred[25,:]).numpy()
\end{verbatim}
The output (within numerical accuracy) is
\begin{verbatim}
    array([9.9911076e-01, 8.8874140e-04, 4.8240952e-07, 5.9778327e-23,
    3.2857954e-15, 2.3312916e-13, 1.7406914e-15, 1.2611157e-24,
    3.6346924e-19, 1.1976714e-26, 2.0100268e-26, 1.2437883e-15,
    1.7560412e-26, 6.3181470e-28, 2.0884368e-21], dtype=float32)
\end{verbatim}
Each value of this array represents the probability that the pattern is from one of the 15 classes. In this case the model predicts with a probability  close to one that the first pattern belongs to the first class $\alpha$ which is correct. When referring to the prediction of the model we typically refer to the class with the largest probability. To evaluate the training history and prediction accuracy of all patterns by a confusion matrix the following script can be used:
\begin{verbatim}
	python CNN_Evaluate.py model_CNN_2D
\end{verbatim}
The following script displays the saliency maps for selected patterns from the training dataset, using model\_CNN\_2D by default:
\begin{verbatim}
	python CNN_Saliency.py
\end{verbatim}

To generate patterns from a dense scan of the parameter space, we use
\begin{verbatim}
	python Parameter_Space_Dataset_Generate.py
\end{verbatim}
This script takes about 24 hours to run. The raw data is also available for download from an OSF repository at Ref.~\onlinecite{Scholz2021b}. To classify the results using the 2D convolutional CNN, run the following command:
\begin{verbatim}
	python Parameter_Space_Dataset_Classify.py model_CNN_2D
\end{verbatim}
To plot all patterns that belong to the same predicted class, we use
\begin{verbatim}
	python Predicted_Class_Montage.py model_CNN_2D 0
\end{verbatim}
To specify the displayed class set the second input parameter to any number from 0 to 14. Here 0 corresponds to class $\alpha$. Clicking on a pattern will output parameters $(k,f)$ to the command prompt.

To generate training and validation data for a second training pass with additional parameters run
\begin{verbatim}
	python Dataset_Generate.py train2 2D
	python Dataset_Generate.py val2 2D
\end{verbatim}
Both scripts run for about 5 hours. The raw data is also available for download from an OSF repository at Ref.~\onlinecite{Scholz2021b}. With the datasets from previous run the second iteration of the training, run
\begin{verbatim}
	python CNN_Train_2D_2nd.py
\end{verbatim}
The script runs for about 1.4 hours. Pre-trained weights are also stored in Ref.~\onlinecite{Scholz2021}. To display results of the new model type
\begin{verbatim}
	python Parameter_Space_Dataset_Classify.py model_CNN_2D_2nd
\end{verbatim}

\end{document}